\documentclass[conference,a4paper]{APSIPA2026}
\usepackage{bm}
\usepackage{dsfont}
\usepackage{amssymb} 
\usepackage{bbm}
\usepackage{algorithm}
\usepackage{algorithmic}
\usepackage{xcolor}
\usepackage{amssymb}
\usepackage{colortbl}
\usepackage{booktabs}
\usepackage{tabularx}
\usepackage{multirow}
\usepackage[flushleft]{threeparttable}
\usepackage{pifont} 
\usepackage[font=small,skip=2pt]{caption}
\usepackage{tcolorbox}

\usepackage{amsmath}
\usepackage{graphicx}
\usepackage{multirow}
\usepackage{threeparttable}
\usepackage{comment}
\usepackage{threeparttable}
\usepackage{xurl} 
\usepackage{hyperref}
\usepackage{booktabs}
\usepackage[backend=biber,style=ieee,]{biblatex}
\usepackage{geometry}
\usepackage{fancyhdr}

\fancypagestyle{firststyle}{
  \fancyhf{}
  \fancyhead[C]{2026 Asia Pacific Signal and Information Processing Association Annual Summit and Conference (APSIPA ASC)}
}

\begin{document}

\title{Unsupervised Single-Channel Speech Separation \\ with Diffusion under Speaker-Embedding Guidance}

\author{
\authorblockN{
Runwu Shi\authorrefmark{1},
Kai Li\authorrefmark{2},
Yiyan Wang\authorrefmark{3},
Jiang Wang\authorrefmark{1},
Chang Li\authorrefmark{4}, 
Ragib Amin Nihal\authorrefmark{1}, 
Sihan Tan\authorrefmark{1},
Kazuhiro Nakadai\authorrefmark{1}
}

\authorblockA{
\authorrefmark{1}
Institute of Science Tokyo,
\authorrefmark{2}
Tsinghua University,\\
\authorrefmark{3}
The University of Tokyo,
\authorrefmark{4}
University of Science and Technology of China
\\
E-mail: shirunwu@ra.sc.e.titech.ac.jp}
}


\maketitle
\thispagestyle{firststyle}
\pagestyle{empty}

\begin{abstract}
Speech separation is a fundamental task in audio processing, typically addressed with fully supervised systems trained on paired mixtures. While effective, such systems typically rely on synthetic data pipelines, which may not reflect real-world conditions. Instead, we revisit the source-model paradigm, training a diffusion generative model solely on anechoic speech and formulating separation as a diffusion inverse problem. However, unconditional diffusion models lack speaker-level conditioning, they can capture local acoustic structure but produce temporally inconsistent speaker identities in separated sources. To address this limitation, we propose Speaker-Embedding guidance that, during the reverse diffusion process, maintains speaker coherence within each separated track while driving embeddings of different speakers further apart. In addition, we propose a new separation-oriented solver tailored for speech separation, and both strategies effectively enhance performance on the challenging task of unsupervised source-model-based speech separation, as confirmed by extensive experimental results. Audio samples and code are available at \href{https://runwushi.github.io/UnSepDiff_demo}{\text{https://runwushi.github.io/UnSepDiff\_demo}}.
\end{abstract}

\section{Introduction}
\label{sec:intro}
\begin{figure}[!tb]
  \centering
  \includegraphics[width=8.2cm]{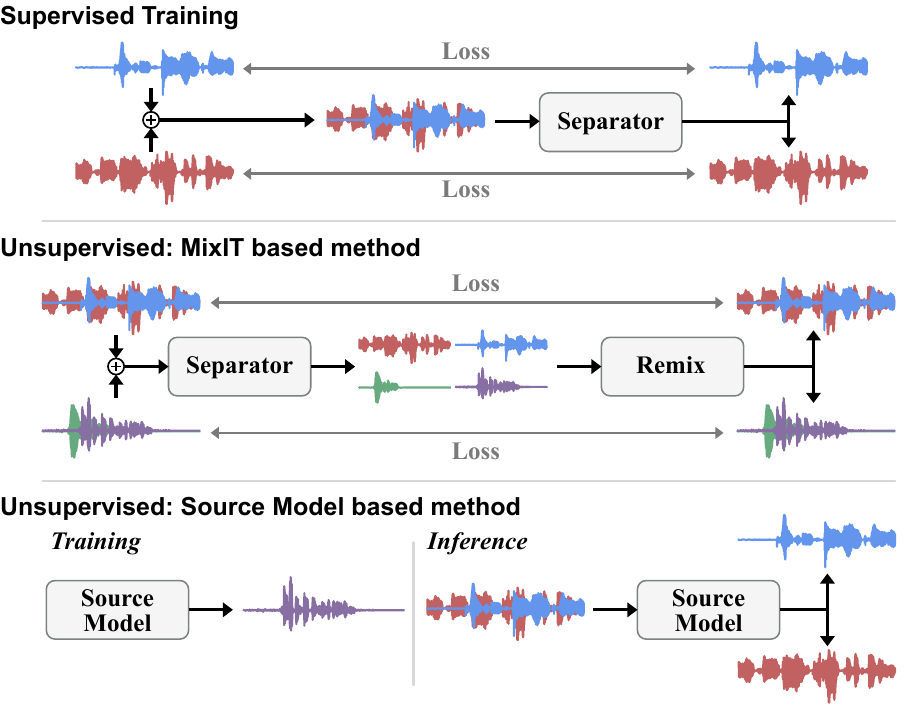}
  \caption{Representative paradigms for single-channel speech separation.}
  \label{fig:1}
\vspace{-1em} 
\end{figure}

Single-channel speech separation aims to recover each speaker's speech signal from a single-channel mixture~\cite{luo2018tasnet, yip2024speech}. As summarized in Figure~\ref{fig:1}, existing approaches can be broadly categorized into three paradigms: supervised training on mixture--source pairs, unsupervised MixIT-style training on mixtures, and source model-based methods. Among them, supervised end-to-end training on synthetic mixture--source pairs has become a dominant and high-performing paradigm~\cite{wang2018supervised, luo2019conv}. This paradigm typically adopts Permutation Invariant Training (PIT)~\cite{yu2017permutation} to address the output ambiguity caused by arbitrary source ordering.

Unsupervised learning has also been explored to reduce the reliance on paired mixture--source data. MixIT proposes an unsupervised framework~\cite{wisdom2020unsupervised} that leverages mixtures of mixtures, enabling separator training directly from mixtures without isolated source signals. However, MixIT may sometimes produce more sources than actually exist, which has motivated subsequent studies to mitigate this issue~\cite{saijo2023self, tzinis2022remixit}. Another unsupervised paradigm is source model-based separation~\cite{benaroya2005audio}, which differs from MixIT-style methods. These approaches first train source models on clean single-source data to capture the intrinsic characteristics of each target source type. During inference, the learned source prior is then used to guide the estimation of the most likely source signals from the observed mixture.

Classical source model-based methods include HMMs \cite{wang2014informed}, NMF \cite{wood2017blind}, Bayesian models \cite{benaroya2005audio}, and VQ-based codebooks \cite{ellis2006model}, while deep generative models such as VAEs \cite{karamatli2019audio}, Glow \cite{zhu2022music}, GANs \cite{narayanaswamy2020unsupervised}, and diffusion models \cite{mariani2023multi} have also been explored. These methods are generally more effective when sources belong to distinct acoustic classes (e.g., drums and piano), where inter-class spectral and temporal differences facilitate the separation \cite{narayanaswamy2020unsupervised}. In contrast, speech separation involves homogeneous sources: all signals belong to the same class, speech, and often exhibit highly overlapping spectral patterns. This similarity makes source model–based methods much more challenging, as the learned prior captures general speech features but lacks discriminative ability to separate individual speakers, particularly when voices are similar (e.g., same gender). Traditional approaches have therefore always focused on speaker-dependent or gender-based settings \cite{bouvier2016source}. The difficulty also persists with modern diffusion-based priors. For instance, \cite{iashchenko2023undiff} employs an unconditional speech diffusion model with analytical sampling but still suffers from speaker ambiguity, in which local segments may be separated, yet consistent assignment across time often fails.

To address the speaker alignment problem, we build on an unconditional diffusion model trained on clean speech as the source prior and introduce a \textbf{Speaker-Embedding Guidance} strategy. During the reverse diffusion process, this guidance encourages intermediate samples of different separated tracks to diverge in the speaker embedding space while maintaining consistency within each track. Experimental results demonstrate that this strategy effectively improves separation performance. In addition, we evaluate representative frameworks for addressing diffusion inverse problems in speech separation, an aspect that has been largely overlooked in prior work.

\section{Background}
We briefly introduce diffusion models and two representative diffusion-based sampling methods for audio separation, which form the foundation of our proposed method.
\subsection{Score-based Diffusion Models}
Score-based diffusion models learn speech priors through a stochastic forward process that corrupts a clean signal $\bm{x}_0 \in \mathbb{R}^D$ to noise over time $t \in [0, T]$, where $D$ is the signal length:
\begin{equation}
    \mathrm{d}\bm{x}= -\frac{\beta(t)}{2}\,\bm{x}\,\mathrm{d}t + \sqrt{\beta(t)}\,\mathrm{d}\bm{w},
\end{equation}
where $\beta(t)$ is the noise schedule and $\bm{w}$ is the Wiener process.
The corresponding reverse process is defined as:
\begin{equation}
    \mathrm{d}\bm{x} = \left[-\frac{\beta(t)}{2}\bm{x} - \beta(t){\nabla_{\bm{x}_t}}\log
    p_{t}(\bm{x}_t)\right]\mathrm{d}t + \sqrt{\beta(t)}\mathrm{d}\bm{\bar{w}},
\end{equation}
where $\bar{\bm{w}}$ is the Wiener process in backward \cite{song2020score}. The score function ${\nabla_{\bm{x}_t}}\log p_{t}(\bm{x}_t)$ is parametrically represented by a neural network $s_{\theta}$, trained via denoising objectives. 

\subsection{Diffusion Posterior Sampling for Audio Separation}
This method is a specific case using Diffusion Posterior Sampling (DPS) for the audio separation task \cite{chung2023diffusion}. In detail, the separation problem can be formulated as $\bm{y}_0 =\;\sum_{k=1}^{K}\bm{x}_0^k + \bm n$, where $\bm{y}_0$ is the observed mixture containing $K$ sources at diffusion timestep 0, and $\bm{n}$ is measurement noise. The goal of recovering sources corresponds to sampling from the posterior probability
\(p(\bm{x}_0^{1:K}|\bm{y}_0)\). This is realized by guiding the reverse diffusion process~(2) with the conditional score \(\nabla_{ \bm{x}_t^k}\log p(\bm{x}_t^k|\bm{y}_0)\), which is decomposed via Bayes’ rule:
\begin{equation}
\nabla_{\bm{x}_t^k} \log p(\bm{x}_t^k | \bm{y}_0) = 
\nabla_{\bm{x}_t^k} \log p(\bm{x}_t^k) + \nabla_{\bm{x}_t^k} \log p(\bm{y}_0|\bm{x}_t^k),
\end{equation}
where $\bm{x}_{t}^{k}$ is the state of $k$th source at diffusion timestep $t$. $\nabla_{\bm{x}_t^k} \log p(\bm{x}_t^k)$ is provided by the learned source model, but the gradient of the liklihood $\nabla_{\bm{x}_t^k} \log p(\bm{y}_0|\bm{x}_t^k)$ is analytically intractable, and DPS proposes to approximate it by $\nabla_{\bm{x}_t^k} \log p(\bm{y}_0|\hat{\bm{x}}_0^k)$, in which $\hat{\bm{x}}_0^k$ is the estimated clean sample via Tweedie's formula. The likelihood gradient of the $ k$th source can then be approximated through backpropagation:
\begin{equation}
\nabla_{\bm{x}_{t}^k} \log p(\bm{y}_0|\bm{x}_{t}^k)
\;\simeq\;
- \gamma(t)\,
\nabla_{\bm{x}_{t}^k} \rVert\bm{y}_0 - \sum_{k=1}^{K} \hat{\bm{x}}_{0}^k\rVert_2^2,
\end{equation}
where $\gamma(t)$ controls guidance strength. DPS uses a constant, while DSG proves noise-proportional $\gamma(t)$ works better \cite{yang2024guidance}.

\subsection{Dirac Sampling for Audio Separation}
\cite{mariani2023multi} proposes a posterior sampling method specific to music separation. In detail, the likelihood $p(\bm{y}_t|\bm{x}_t^k)$ at diffusion timestep $t$ is approximated by a Dirac function $p(\bm{y}_t|\bm{x}_t^k)=\mathbbm{1}_{\bm{y}_t=\sum_{k=1}^{K} \bm{x}_t^k}$, which directly utilize the summation of sources at different diffusion timestep, and when sources are independent, the corresponding posterior score is defined as:
\begin{equation}
\nabla_{\bm{x}_t^k} \log p(\bm{x}_t^k | \bm{y}_0) \simeq 
s_{\theta}(\bm{x}_t^k,t)-\xi(t)(\bm{y}_0-\sum_{k=1}^{K}\bm{x}_t^k)
,
\end{equation}
where $\xi(t)$ is a tunable coefficient, $s_{\theta}$ is the learned diffusion source model. This method provides tight conditioning without the need for backpropagation, and we find this strategy is pretty effective for unsupervised speech separation.

\section{Methodology}
This section introduces the proposed Speaker-Embedding guided speech separation solver, which integrates Dirac Sampling and DPS-based strategies. 
During the first stage of the reverse diffusion process, we employ Dirac Sampling combined with Speaker-Embedding guidance. 
The Dirac Sampling provides a stable separation, while the speaker guidance in this regime can thus influence the assignment of speaker identities across different separated tracks, facilitating the discovery of sources. 
In the second stage, we switch to a backpropagation-based DPS-style solver, which enables higher-quality signal reconstruction. The overall sampling process of the separation is detailed in algorithm \ref{alg:1}.

\begin{figure*}[!tb]
  \centering
  \includegraphics[width=14.5cm]{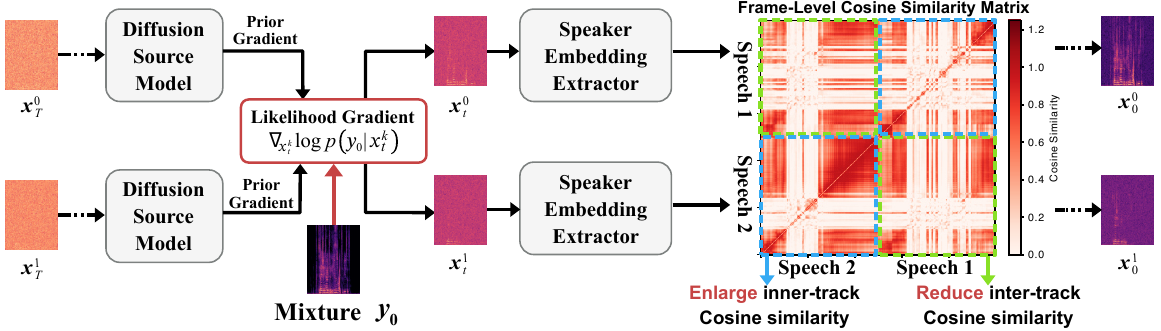}
  \caption{Speaker-Embedding Guided Speech Separation. The likelihood gradient can be obtained through DPS, Dirac sampling, etc.}
  \label{fig:2}
\end{figure*}

\subsection{Speaker-Embedding Guided Speech Separation}
To address the temporal inconsistency of speaker identities inherent in unconditional source models, we introduce a Speaker-Embedding guidance mechanism, as shown in Figure \ref{fig:2}. This mechanism leverages a pre-trained speaker embedding extractor $\mathcal{E}_{\text{spk}}$, which maps a speech segment into a compact representation of speaker identity. Our core idea is to utilize this representation to steer the reverse diffusion trajectory toward solutions that preserve consistent speaker identities. For the $k$ th separated track intermediate state $\bm{x}_t^k$, we compute its speaker embedding $\mathbf{E}_{\text{spk}}^k = \mathcal{E}_{\text{spk}}(\bm{x}_t^k)$. Next, we adopt the guidance step following the principle of classifier guidance \cite{dhariwal2021diffusion}, but instead of aligning with a predefined class label, the objective here is to enforce speaker-level constraints. Specifically, we formulate a speaker loss $\mathcal{L}_{\text{spk}}$ that encourages both \textbf{inner-track consistency} and \textbf{inter-track distinctiveness}. Inner-track consistency is enforced by maximizing the cosine similarity between embeddings along the temporal dimension within the same track, thereby ensuring stable speaker identity over time. Inter-track distinctiveness is enforced by minimizing the cosine similarity between embeddings from different separated tracks. The combined loss $\mathcal{L}_{\text{spk}}$ is given by:  
\begin{equation}
\mathcal{L}_{\text{spk}}
= \sum_k \big( 1 - \cos(\mathbf{E}^k_\text{spk}, \mathbf{E}^k_\text{spk}) \big)
+ \sum_{k \neq j} \cos(\mathbf{E}^k_\text{spk}, \mathbf{E}^j_\text{spk}),
\end{equation}

We then backpropagate $\mathcal{L}_{\text{spk}}$ with respect to $\{ \bm{x}_t^k \}_{k=1}^{K}$ to obtain the gradient, normalize it and use this direction to steer the update of $\bm{x}_t^k$ during the reverse diffusion process:
\begin{equation}
\nabla_{\bm{x}_t^k}\log p(\bm{y}|\bm{x}_t^k) \approx 
\nabla_{\bm{x}_t^k}\mathcal{L}_{\text{spk}} \,/\, \|\nabla_{\bm{x}_t^k}\mathcal{L}_{\text{spk}}\|_2,
\end{equation}
\begin{equation}
\bm{x}_{t}^k \;\leftarrow\; \bm{x}_{t}^{k} - \nabla_{\bm{x}_t^k} \log p(\bm{y}|\bm{x}_t^k).
\end{equation}

This speaker-embedding guidance process can be a flexible plug-in, applicable to various diffusion-based separation solvers like Dirac sampling and DPS. Specifically, we employ Dirac sampling in the first stage $[0, T_{\text{dirac}}]$, and switch to DPS-based updates in the second stage for $T_D$ iterations. The speaker-embedding guidance is applied within a interval $[T_{\text{spk}}^{\text{start}}, T_{\text{spk}}^{\text{end}}]$ within the Dirac sampling process. 

\begin{algorithm}[t]
\caption{Speaker-Embedding Guided Speech Separation}
\label{alg:1}
\textbf{Require:} Diffusion steps $T$, DPS steps $T_D$, schedule $\{\sigma_t\}$, $\{\sigma^{\text{post}}_t\}$, source model $s_\theta$, speaker model $\mathcal{E}_{\text{spk}}$, dimension $D$ \\
\textbf{Input:} mixture $\bm{y}$, $K$ sources
\begin{algorithmic}[1]
\STATE Initialize $\{\bm{x}_T^k\}_{k=1}^K \sim \mathcal{N}(0,\mathbf{I})$ \hfill // initialization
\FOR{$t = T$ \TO $1$}

\FOR{$k = 1$ \TO $K$}
  \IF{$k=a$} 
    \STATE $\bm{x}_t^{a} \leftarrow \bm{y} - \sum_{\substack{j=1 \\ j\neq a}}^K \bm{x}_t^j$ \hfill // Dirac Sampling, $a$: anchor 
  \ENDIF
  \STATE $\text{score}_t^k \leftarrow S_\theta^k(\bm{x}_t^k,\sigma_t)$ \hfill // model score
\ENDFOR

\colorbox{red!8}{%
  \parbox{\dimexpr\linewidth-2\fboxsep}{%
    \STATE {\textit{\textbf{Speaker-Embedding Guidance}:}}
    \IF{$t \in [T_{\text{spk}}^{\text{start}}, T_{\text{spk}}^{\text{end}}]$}
      \STATE $ \{\mathbf{E}_{\text{spk}}^k \} \leftarrow \mathcal{E}_{\text{spk}}(\{\bm{x}_t^k\})$ \hfill // extract spk. embeddings
      \STATE $\bm{g} \leftarrow \nabla_{\{\bm{x}_t^k\}} \mathcal{L}_{\text{spk}}$, \hfill $r \leftarrow \sqrt{D}\cdot \sigma^{\text{post}}_t$  \; \hfill // gradient \;
      \STATE $\text{score}_t^k \leftarrow \text{score}_t^k - r \cdot {\bm{g}}/{\|\bm{g}\|_2}$ \hfill // score update
    \ENDIF
  }%
}

\STATE $\{\Delta\text{score}_t^k\}_{k=1}^K \leftarrow \{\text{score}_t^k - \text{score}_t^a\}_{k=1}^K$
\STATE $\{\bm{x}_{t}^k\}_{k=1}^K \leftarrow 
       \{\bm{x}_{t}^k + (\sigma_{t+1} - \hat\sigma_t)\,\Delta\text{score}_t^k\}_{k=1}^K$
\STATE $\{\bm{x}_{t}^k\}_{k=1}^K \leftarrow 
\{\bm{x}_{t}^k + \sqrt{\sigma_t^{2}-\hat\sigma_t^{2}}\,\bm{\epsilon}_t^k\}_{k=1}^K$, $\bm{\epsilon}_t^k \sim \mathcal{N}(0,\mathbf{I})$ 
\ENDFOR
\FOR{$t = T_D$ \TO 1} 
\STATE $\Big\{ \bm{x}_{t-1}^k \Big\}_{k=1}^K 
\leftarrow 
\Big\{ \bm{x}_{t-1}^{k'} - 
\nabla_{\bm{x}_t^k}\big\|\bm{y}-\sum_{j=1}^K \hat{\bm{x}}_0^j\big\|_2^2 
\Big\}_{k=1}^K$
\ENDFOR 
\hfill // DPS refinement and update
\STATE \textbf{return} $\{\bm{x}_{0}^{\,k}\}_{k=1}^K$
\end{algorithmic}
\end{algorithm}

\subsection{Speaker-Embedding Extractor}
We design the speaker-embedding extractor as a cascaded architecture, where a denoise module is followed by a speaker encoder network, since the extractor should work in noisy diffusion space. The denoise module GTCRN \cite{rong2024gtcrn} first suppresses interference in $\bm{x}_t^k$ to enhance noise robustness. Its output is then converted into mel-spectrograms and passed to the speaker encoder NeXt-TDNN \cite{heo2024next}, which produces frame-level speaker embeddings.

\begin{table*}[t]
  \caption{Comparison on 2-second VCTK-2mix and mismatched WSJ0-2mix. Unsupervised bests are \underline{underlined}, overall bests are in \textbf{bold}.}
  \label{tab:1}
  \centering
  \small
\begin{threeparttable}
\begin{tabularx}{\textwidth}{
  >{\raggedright\arraybackslash}p{4.0cm} |
  >{\centering\arraybackslash}p{1.7cm} |
  >{\centering\arraybackslash}p{1.5cm}
  >{\centering\arraybackslash}p{1.2cm}
  >{\centering\arraybackslash}p{1.5cm} |
  >{\centering\arraybackslash}p{1.5cm}
  >{\centering\arraybackslash}p{1.2cm}
  >{\centering\arraybackslash}p{1.5cm}}
  \multicolumn{2}{c}{} & \multicolumn{3}{c}{\textbf{VCTK-2mix (Trained)}} & \multicolumn{3}{c}{\textbf{WSJ0-2mix (Mismatched)}} \\
  \toprule
  \multicolumn{1}{c|}{\textbf{Method}} & \textbf{Supervised} 
    & \textbf{SI-SDR}$\uparrow$ & \textbf{SDR}$\uparrow$ & \textbf{STOI}$\uparrow$
    & \textbf{SI-SDR}$\uparrow$ & \textbf{SDR}$\uparrow$ & \textbf{STOI}$\uparrow$ \\
  \midrule
    Unprocessed & -- & 0.003 & 0.000 & 0.73 & 0.003 & 0.000 & 0.70 \\
    \midrule
    Conv-TasNet \cite{luo2019conv} & {\large \ding{51}} & 12.00 & 12.56 & 0.86 & 5.10 & 6.67 & 0.82 \\
    TF-Locoformer \cite{saijo2024tf} & {\large \ding{51}} & \textbf{15.32} & \textbf{15.61} & \textbf{0.90} & \textbf{10.11} & \textbf{10.95} & \textbf{0.88} \\
    \midrule
    DPS \cite{chung2023diffusion} & {\large \ding{55}} & -3.43 & 0.28 & 0.68 & -11.76 & 0.20 & 0.58 \\
    DSG \cite{yang2024guidance} & {\large \ding{55}} & 4.99 & 6.41 & 0.73 & 0.81 & 4.09 & 0.68 \\
    Analytical Sampling \cite{iashchenko2023undiff} & {\large \ding{55}} & 2.45 & 3.64 & 0.75 & 1.54 & 3.78 & 0.76 \\
    Dirac Sampling \cite{mariani2023multi} & {\large \ding{55}} & 8.41 & 9.39 & \underline{\textbf{0.82}} & 4.15 & 5.68 & 0.79 \\
    \midrule
    Proposed Sampling & {\large \ding{55}} & 8.46 & 9.36 & 0.80 & 4.63 & 5.98 & \underline{\textbf{0.80}} \\
    Proposed Sampling + Speaker & {\large \ding{55}} & \underline{\textbf{9.32}} & \underline{\textbf{10.33}} & \underline{\textbf{0.82}} & \underline{\textbf{4.79}} & \underline{\textbf{6.10}} & \underline{\textbf{0.80}}\\
    \bottomrule
  \end{tabularx}
  \end{threeparttable}
\end{table*}

\section{Experiments and Results}
\subsection{Dataset and Model Configuration}
We adopt the VCTK corpus \cite{yamagishi2019cstr} consisting of 110 speakers, using 100 speakers for training the diffusion source model and the remaining 10 for evaluation. To assess out-of-domain generalization, we further test on the WSJ0-2mix dataset. As the source prior, we employ a frequency-domain diffusion model that operates in the waveform space by directly predicting complex-valued STFT coefficients. The model architecture follows a DPRNN-style design with temporal and spectral self-attention~\cite{luo2020dual}, augmented by a global temporal attention module in the latent layer. The resulting model has 37M parameters and is trained on VCTK resampled to 16 kHz for 400k steps using AdamW \cite{loshchilov2017decoupled}, with a linear noise schedule where $T=200$ and $\beta \in [10^{-4}, 2\times10^{-2}]$. The implementation and pretrained checkpoint are publicly available \footnote{{https://github.com/RunwuShi/UnSepDiff}}. For the speaker-embedding extractor, we train a network on the VCTK dataset, using the same 100 training speakers as the diffusion model. The model is trained for 400 epochs with a joint objective combining softmax-based speaker classification and a SI-SDR loss to improve noise robustness. It operates on 80-bin mel-spectrogram features. For scheduling, the Dirac sampling step is fixed at $T_{\text{dirac}}=200$, while the speaker-guidance interval is set to $T_{\text{spk}}^{\text{start}}=75$ and $T_{\text{spk}}^{\text{end}}=175$ within the Dirac sampling stage. We adopt a single-step DPS with $T_D=1$ after the Dirac sampling to balance consistency and quality, as excessive DPS tends to increase speaker inconsistency.

\subsection{Experimental Setup and Baselines}
For in-domain evaluation, we synthesize 400 mixtures from the VCTK dataset by randomly overlapping single-source utterances from unseen speakers. Both the Root Mean Square (RMS) levels (sampled uniformly between –25 and –20 dB) and the temporal offsets are randomized to increase diversity. For out-of-domain evaluation, we randomly select 400 mixtures from the WSJ0-2mix test set. We compare our method against supervised baselines and diffusion-based sampling approaches. As supervised baselines, we adopt the time-domain Conv-TasNet~\cite{luo2019conv} and the frequency-domain TF-Locoformer small~\cite{saijo2024tf}, each trained for 200 epochs on the VCTK dataset with dynamic mixing, and each epoch contains 2,000 samples. Both models are trained using PIT with SI-SDR loss. For unsupervised diffusion-based baselines, we include DPS~\cite{chung2023diffusion} and DSG~\cite{yang2024guidance}, which serve as advanced generic diffusion inverse solvers, as well as Analytic Sampling~\cite{iashchenko2023undiff} and Dirac Sampling~\cite{mariani2023multi}, which are specifically designed for audio separation. We evaluate the proposed hybrid sampling method without and with Speaker-Embedding guidance, reported as Proposed Sampling and Proposed Sampling + Speaker.

 \begin{table*}[t]
      \centering
\caption{
We report the ratio of samples where both separated sources achieve positive SI-SDR, the corresponding successful SI-SDR (Succ. SI-SDR), and three speaker-similarity metrics computed over the full test set.
Higher is better for all metrics except mean cross-speaker cosine similarity. The best results are marked with \underline{\textbf{bold underline}}, and the second-best results are marked in \textbf{bold}.}
      \label{tab:tse_spk_similarity}
      \small
      \begin{tabular}{lccccc}
      \toprule
      Method
      & SI-SDR $> 0$ Ratio $\uparrow$
      & Succ. SI-SDR $\uparrow$
      & Spk. Sim. $\uparrow$
      & Cross Spk. Sim. $\downarrow$
      & Sim. Margin $\uparrow$ \\
      \midrule
      DPS \cite{chung2023diffusion}
      & 41.00\%
      & 6.59
      & \textbf{0.537}
      & 0.257
      & 0.279 \\
      DSG \cite{yang2024guidance}
      & 70.75\%
      & 8.13
      & \underline{\textbf{0.594}}
      & 0.227
      & \underline{\textbf{0.367}}\\
      Analytical Sampling \cite{iashchenko2023undiff}
      & 61.50\%
      & 5.32
      & 0.518
      & 0.294
      & 0.225 \\
      Dirac Sampling \cite{mariani2023multi}
      & 81.00\%
      & \textbf{11.02}
      & 0.421
      & \textbf{0.136}
      & 0.286 \\
      \midrule
      Proposed Sampling
      & \textbf{81.50\%}
      & 10.94
      & 0.345
      & \underline{\textbf{0.114}}
      & 0.230 \\
      Proposed Sampling + Speaker
      & \underline{\textbf{81.75\%}}
      & \underline{\textbf{11.94}}
      & {{0.441}}
      & {{0.143}}
      & \textbf{0.298} \\
      \bottomrule
      \end{tabular}
  \end{table*}

\subsection{Results and Discussion}
We evaluate our method on both the VCTK-2mix and the mismatched WSJ0-2mix datasets, focusing on 2-second mixtures. We report results using SI-SDR, SDR, and Short-Time Objective Intelligibility (STOI) as evaluation metrics. As shown in Table~\ref{tab:1}, the supervised TF-Locoformer~\cite{saijo2024tf} achieves the best overall performance across all metrics. Among unsupervised baselines, our proposed sampling with Speaker-Embedding guidance achieves the strongest unsupervised results across all metrics, both on the in-domain VCTK dataset and on the out-of-domain WSJ0-2mix dataset. Even without speaker embeddings, the proposed sampling improves over most unsupervised baselines, yielding higher SI-SDR on VCTK and consistently better results on WSJ0-2mix. Notably, the WSJ0-2mix benchmark involves a larger and more diverse set of speakers, which makes the separation task considerably more challenging and accounts for the overall lower performance across methods. While supervised systems still deliver the highest absolute scores, these findings confirm that providing an unconditional diffusion source model with additional Speaker-Embedding guidance effectively enhances separation performance, highlighting the central contribution of our approach.

Moreover, we visualize the per-sample SI-SDR distributions of different unsupervised diffusion-based sampling strategies, as shown in Fig.~\ref{fig:2}. Each dot corresponds to one separated mixture, where the SI-SDR is averaged over the two separated sources after permutation matching. The black circle denotes the mean SI-SDR over all evaluated mixtures for each method. Since our proposed solver is mainly built upon the Dirac sampling strategy, its overall distribution follows a similar pattern to Dirac. However, by introducing Speaker-Embedding guidance, the proposed method further improves the upper range of the distribution, leading to more samples achieving high SI-SDR values. This indicates that the speaker-level information provides an effective additional constraint for the diffusion reverse process, helping the model produce more speaker-consistent separated sources and correct some possible assignment errors. As a result, the proposed guidance raises the average performance over the Dirac baseline, demonstrating the benefit of incorporating speaker-discriminative information into source-model-based unsupervised speech separation.

\begin{figure}[!tb]
  \centering
  \includegraphics[width=8cm]{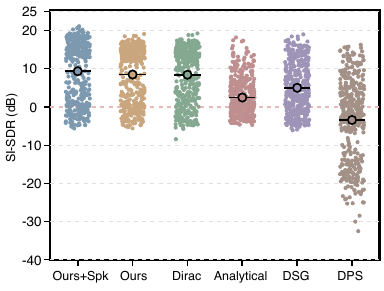}
  \caption{Per-sample SI-SDR comparison across methods on 400 mixture samples from the VCTK-2mix test dataset. }
  \label{fig:2}
\end{figure}

Besides separation quality, we further evaluate speaker consistency using a pretrained NeXt-TDNN speaker encoder~\cite{heo2024next}, which is trained on clean speech for 400 epochs. The results are shown in Table~\ref{tab:tse_spk_similarity}. For each separated mixture, we first compute the SI-SDR of the two estimated sources after permutation matching. We then report the ratio of samples for which both separated sources obtain positive SI-SDR, denoted as SI-SDR $>0$ Ratio. The corresponding Succ. SI-SDR is computed by averaging the SI-SDR over these successful samples. This metric reflects the separation quality when the method produces non-degraded estimates for both sources.

To further analyze speaker assignment, we compute the cosine similarity between each separated signal and each reference source using the pretrained speaker encoder. We report three speaker-similarity metrics over the full test set: Spk. Sim., Cross Spk. Sim., and Sim. Margin. Spk. Sim. denotes the cosine similarity under the correct source-speaker matching, while Cross Spk. Sim. denotes the similarity to the non-matching speaker. Sim. Margin is defined as the difference between these two scores, measuring how clearly the separated output is assigned to the correct speaker. Therefore, higher Spk. Sim. and Sim. Margin indicate better speaker consistency, while lower Cross Spk. Sim. is preferred.

As shown in Table~\ref{tab:tse_spk_similarity}, Dirac Sampling already achieves a much higher SI-SDR $>0$ Ratio and Succ. SI-SDR than DPS, DSG, and Analytical Sampling, indicating that it provides a stronger base solver for source-model-based speech separation. Our proposed sampling further preserves this advantage. Although Proposed Sampling without speaker guidance slightly decreases the speaker-similarity scores compared with Dirac Sampling, this is mainly because the additional DPS-style correction improves local signal clarity but may disturb the global source-speaker assignment. After introducing Speaker-Embedding guidance, both separation quality and speaker consistency are improved: Proposed Sampling + Speaker achieves the best SI-SDR $>0$ Ratio and Succ. SI-SDR, while also increasing Spk. Sim. and Sim. Margin compared with Proposed Sampling. This confirms that speaker guidance helps compensate for the assignment ambiguity introduced during the reverse diffusion process.

We note that the absolute speaker-similarity values are relatively moderate. This is because the speaker encoder is trained on clean speech, whereas diffusion-based separated outputs may still contain residual artifacts or low-level noise, leading to a domain mismatch in speaker embedding extraction. In addition, different methods exhibit a trade-off between perceptual clarity and speaker assignment. DPS and DSG can sometimes produce clearer local speech segments, resulting in higher speaker similarity scores, but they are less reliable in producing correct source assignments and positive SI-SDR for both sources. In contrast, Dirac-based methods, including ours, are better at recovering correctly paired sources, while the outputs may be less clean in some cases. The proposed Speaker-Embedding guidance alleviates this trade-off by improving the high-quality successful cases and enhancing speaker consistency, which is also supported by the listening examples on our demo page.
  
\section{Conclusion}
In this work, we revisited source-model-based speech separation and showed that diffusion models trained on clean speech can be adapted to unsupervised speech separation. To address the lack of speaker conditioning, we proposed Speaker-Embedding guidance to maintain temporal coherence and enlarge speaker separability, together with a separation-oriented solver to improve separation quality. These strategies yield consistent improvements in unsupervised diffusion-based separation.

\printbibliography

@inproceedings{luo2018tasnet,
  title={Tasnet: time-domain audio separation network for real-time, single-channel speech separation},
  author={Luo, Yi and Mesgarani, Nima},
  booktitle={2018 IEEE International Conference on Acoustics, Speech and Signal Processing (ICASSP)},
  pages={696--700},
  year={2018},
  organization={IEEE}
}

@article{wang2018supervised,
  title={Supervised speech separation based on deep learning: An overview},
  author={Wang, DeLiang and Chen, Jitong},
  journal={IEEE/ACM transactions on audio, speech, and language processing},
  volume={26},
  number={10},
  pages={1702--1726},
  year={2018},
  publisher={IEEE}
}

@article{luo2019conv,
  title={Conv-tasnet: Surpassing ideal time--frequency magnitude masking for speech separation},
  author={Luo, Yi and Mesgarani, Nima},
  journal={IEEE/ACM transactions on audio, speech, and language processing},
  volume={27},
  number={8},
  pages={1256--1266},
  year={2019},
  publisher={IEEE}
}

@inproceedings{yu2017permutation,
  title={Permutation invariant training of deep models for speaker-independent multi-talker speech separation},
  author={Yu, Dong and Kolb{\ae}k, Morten and Tan, Zheng-Hua and Jensen, Jesper},
  booktitle={2017 IEEE International Conference on Acoustics, Speech and Signal Processing (ICASSP)},
  pages={241--245},
  year={2017},
  organization={IEEE}
}

@article{wisdom2020unsupervised,
  title={Unsupervised sound separation using mixture invariant training},
  author={Wisdom, Scott and Tzinis, Efthymios and Erdogan, Hakan and Weiss, Ron and Wilson, Kevin and Hershey, John},
  journal={Advances in neural information processing systems},
  volume={33},
  pages={3846--3857},
  year={2020}
}

@inproceedings{saijo2023self,
  title={Self-remixing: Unsupervised speech separation via separation and remixing},
  author={Saijo, Kohei and Ogawa, Tetsuji},
  booktitle={ICASSP 2023-2023 IEEE International Conference on Acoustics, Speech and Signal Processing (ICASSP)},
  pages={1--5},
  year={2023},
  organization={IEEE}
}

@article{tzinis2022remixit,
  title={Remixit: Continual self-training of speech enhancement models via bootstrapped remixing},
  author={Tzinis, Efthymios and Adi, Yossi and Ithapu, Vamsi K and Xu, Buye and Smaragdis, Paris and Kumar, Anurag},
  journal={IEEE Journal of Selected Topics in Signal Processing},
  volume={16},
  number={6},
  pages={1329--1341},
  year={2022},
  publisher={IEEE}
}

@article{benaroya2005audio,
  title={Audio source separation with a single sensor},
  author={Benaroya, Laurent and Bimbot, Fr{\'e}d{\'e}ric and Gribonval, R{\'e}mi},
  journal={IEEE Transactions on Audio, Speech, and Language Processing},
  volume={14},
  number={1},
  pages={191--199},
  year={2005},
  publisher={IEEE}
}

@inproceedings{ellis2006model,
  title={Model-based monaural source separation using a vector-quantized phase-vocoder representation},
  author={Ellis, Daniel PW and Weiss, Ron J},
  booktitle={2006 IEEE International Conference on Acoustics Speech and Signal Processing Proceedings},
  volume={5},
  pages={V--V},
  year={2006},
  organization={IEEE}
}

@article{wang2014informed,
  title={Informed single-channel speech separation using HMM--GMM user-generated exemplar source},
  author={Wang, Qi and Woo, Wai Lok and Dlay, Satnam Singh},
  journal={IEEE/ACM Transactions on Audio, Speech, and Language Processing},
  volume={22},
  number={12},
  pages={2087--2100},
  year={2014},
  publisher={IEEE}
}

@article{karamatli2019audio,
  title={Audio source separation using variational autoencoders and weak class supervision},
  author={Karamatl{\i}, Ertu{\u{g}} and Cemgil, Ali Taylan and K{\i}rb{\i}z, Serap},
  journal={IEEE Signal Processing Letters},
  volume={26},
  number={9},
  pages={1349--1353},
  year={2019},
  publisher={IEEE}
}

@inproceedings{narayanaswamy2020unsupervised,
  title={Unsupervised Audio Source Separation Using Generative Priors},
  author={Narayanaswamy, Vivek and Thiagarajan, Jayaraman J and Anirudh, Rushil and Spanias, Andreas},
  booktitle={Proc. Interspeech 2020},
  pages={2657--2661},
  year={2020}
}

@article{zhu2022music,
  title={Music source separation with generative flow},
  author={Zhu, Ge and Darefsky, Jordan and Jiang, Fei and Selitskiy, Anton and Duan, Zhiyao},
  journal={IEEE Signal Processing Letters},
  volume={29},
  pages={2288--2292},
  year={2022},
  publisher={IEEE}
}

@article{mariani2023multi,
  title={Multi-source diffusion models for simultaneous music generation and separation},
  author={Mariani, Giorgio and Tallini, Irene and Postolache, Emilian and Mancusi, Michele and Cosmo, Luca and Rodol{\`a}, Emanuele},
  journal={arXiv preprint arXiv:2302.02257},
  year={2023}
}

@article{wood2017blind,
  title={Blind speech separation and enhancement with GCC-NMF},
  author={Wood, Sean UN and Rouat, Jean and Dupont, St{\'e}phane and Pironkov, Gueorgui},
  journal={IEEE/ACM Transactions on Audio, Speech, and Language Processing},
  volume={25},
  number={4},
  pages={745--755},
  year={2017},
  publisher={IEEE}
}

@inproceedings{bouvier2016source,
  title={A source/filter model with adaptive constraints for NMF-based speech separation},
  author={Bouvier, Damien and Obin, Nicolas and Liuni, Marco and Roebel, Axel},
  booktitle={2016 ieee international conference on acoustics, speech and signal processing (icassp)},
  pages={131--135},
  year={2016},
  organization={IEEE}
}

@inproceedings{iashchenko2023undiff,
  title={UnDiff: Unsupervised Voice Restoration with Unconditional Diffusion Model},
  author={Iashchenko, Anastasiia and Andreev, Pavel and Shchekotov, Ivan and Babaev, Nicholas and Vetrov, Dmitry},
  booktitle={Proc. Interspeech 2023},
  pages={4294--4298},
  year={2023}
}

@article{song2020score,
  title={Score-based generative modeling through stochastic differential equations},
  author={Song, Yang and Sohl-Dickstein, Jascha and Kingma, Diederik P and Kumar, Abhishek and Ermon, Stefano and Poole, Ben},
  journal={arXiv preprint arXiv:2011.13456},
  year={2020}
}

@inproceedings{chung2023diffusion,
  title={Diffusion Posterior Sampling for General Noisy Inverse Problems},
  author={Chung, Hyungjin and Kim, Jeongsol and Mccann, Michael T and Klasky, Marc L and Ye, Jong Chul},
  booktitle={The Eleventh International Conference on Learning Representations, ICLR 2023},
  year={2023},
  organization={The International Conference on Learning Representations}
}

@inproceedings{yang2024guidance,
  title={Guidance with Spherical Gaussian Constraint for Conditional Diffusion},
  author={Yang, Lingxiao and Ding, Shutong and Cai, Yifan and Yu, Jingyi and Wang, Jingya and Shi, Ye},
  booktitle={International Conference on Machine Learning},
  pages={56071--56095},
  year={2024},
  organization={PMLR}
}

@inproceedings{heo2024next,
  title={NeXt-TDNN: Modernizing multi-scale temporal convolution backbone for speaker verification},
  author={Heo, Hyun-Jun and Shin, Ui-Hyeop and Lee, Ran and Cheon, YoungJu and Park, Hyung-Min},
  booktitle={ICASSP 2024-2024 IEEE International Conference on Acoustics, Speech and Signal Processing (ICASSP)},
  pages={11186--11190},
  year={2024},
  organization={IEEE}
}

@inproceedings{rong2024gtcrn,
  title={GTCRN: A speech enhancement model requiring ultralow computational resources},
  author={Rong, Xiaobin and Sun, Tianchi and Zhang, Xu and Hu, Yuxiang and Zhu, Changbao and Lu, Jing},
  booktitle={ICASSP 2024-2024 IEEE International Conference on Acoustics, Speech and Signal Processing (ICASSP)},
  pages={971--975},
  year={2024},
  organization={IEEE}
}

@inproceedings{luo2020dual,
  title={Dual-path rnn: efficient long sequence modeling for time-domain single-channel speech separation},
  author={Luo, Yi and Chen, Zhuo and Yoshioka, Takuya},
  booktitle={ICASSP 2020-2020 IEEE International Conference on Acoustics, Speech and Signal Processing (ICASSP)},
  pages={46--50},
  year={2020},
  organization={IEEE}
}

@inproceedings{saijo2024tf,
  title={TF-Locoformer: Transformer with local modeling by convolution for speech separation and enhancement},
  author={Saijo, Kohei and Wichern, Gordon and Germain, Fran{\c{c}}ois G and Pan, Zexu and Le Roux, Jonathan},
  booktitle={2024 18th International Workshop on Acoustic Signal Enhancement (IWAENC)},
  pages={205--209},
  year={2024},
  organization={IEEE}
}

@article{dhariwal2021diffusion,
  title={Diffusion models beat gans on image synthesis},
  author={Dhariwal, Prafulla and Nichol, Alexander},
  journal={Advances in neural information processing systems},
  volume={34},
  pages={8780--8794},
  year={2021}
}

@article{loshchilov2017decoupled,
  title={Decoupled weight decay regularization},
  author={Loshchilov, Ilya and Hutter, Frank},
  journal={arXiv preprint arXiv:1711.05101},
  year={2017}
}

@inproceedings{yip2024speech,
  title={Speech Separation using Neural Audio Codecs with Embedding Loss},
  author={Yip, Jia Qi and Kwok, Chin Yuen and Ma, Bin and Chng, Eng Siong},
  booktitle={2024 Asia Pacific Signal and Information Processing Association Annual Summit and Conference (APSIPA ASC)},
  pages={1--6},
  year={2024},
  organization={IEEE}
}

@article{yamagishi2019cstr,
  title={CSTR VCTK Corpus: English multi-speaker corpus for CSTR voice cloning toolkit (version 0.92)},
  author={Yamagishi, Junichi and Veaux, Christophe and MacDonald, Kirsten},
  journal={The Rainbow Passage which the speakers read out can be found in the International Dialects of English Archive:(http://web. ku. edu/\~{} idea/readings/rainbow. htm).},
  year={2019},
  publisher={University of Edinburgh. The Centre for Speech Technology Research (CSTR)}
}

\end{document}